\documentstyle[12pt]{article}
\begin{document}

\author{R. Vilela Mendes \\ 
Grupo de F\'\i sica-Matem\'atica\\
Complexo II, Universidade de Lisboa \\
Av. Gama Pinto, 2, 1699 Lisboa Codex Portugal\\
e-mail: vilela@alf4.cii.fc.ul.pt}
\title{Active control of ionized boundary layers}
\date{}
\maketitle

\begin{abstract}
The challenging problems, in the field of control of chaos or of transition
to chaos, lie in the domain of infinite-dimensional systems. Access to all
variables being impossible in this case and the controlling action being
limited to a few collective variables, it will not in general be possible to
drive the whole system to the desired behaviour. A paradigmatic problem of
this type is the control of the transition to turbulence in the boundary
layer of fluid motion. By analysing a boundary layer flow for an ionized
fluid near an airfoil, one concludes that active control of the transition
amounts to the resolution of an generalized integro-differential eigenvalue
problem. To cope with the required response times and phase accuracy,
electromagnetic control, whenever possible, seems more appropriate than
mechanical control by microactuators.
\end{abstract}

\section{Introduction}

Control of chaos or of the transition to chaos has been, in recent years, a
very active field (see for example Ref.\cite{Shinbrot} and references
therein). Several techniques were developed and tested, mostly for low
dimensional dynamical systems. The challenge lies now on finding out whether
these techniques extend to infinite-dimensional systems.

A first aspect preventing a simple extrapolation of the finite-dimensional
techniques is the fact that only a small subset of variables (or some
integrated collective variable) is acessible to measurement. Likewise the
variables on which one may act for controlling purposes are even more
limited. A second aspect is that, rather than to stabilize an unstable
periodic orbit (a single mode), what one aims in general is to suppress a
continuous set of unstable modes, or to stabilize a particular collective
mode and, at the same time, prevent all other modes from developing. In this
sense the problem is no longer a standard control problem to be handled by
pole placement, sliding mode or other standard technique. Instead, as
suggested by the problem discussed in this paper the control problem amounts
to the solution of a generalized integro-differential eigenvalue problem.

A problem of both theoretical and practical importance is the control of the
transition from laminar to turbulent motion in a boundary layer flow. I deal
with this problem mostly as an example and prototype of the kind of
questions and mathematical framework to be expected in the control of chaos
for infinite-dimensional complex systems. However for the benefit of the
reader less familiar with aerodynamical issues I have included a few remarks
on the physical and technological context of the problem.

By delaying the laminar to turbulent transition, an order of magnitude
reduction in the skin friction drag is achieved. The technological benefits
that may be derived from this reduction, led to the proposal of several
methods for the control of the boundary layer transition. They are both of
passive and active type and include pressure gradient control, wall suction,
wall temperature control, polymer coating, compliant walls, etc.

In passive type control\cite{Bushnell} \cite{Vilela}, the aim is either to
induce a modification of the curvature of the velocity profile, or to break
the eddies and absorb their energy.

On the other hand the active control methods, that have been proposed\cite
{Joslin}, aim at cancelling the growth of the Tollmien-Schlichting (TS)
waves, a known precursor of the transition instability. This is achieved by
creating a disturbance of opposite phase to cancel the TS waves. The wave
cancelling disturbance may be created, for example, by modulated suction and
blowing or by mechanical microactuators. This control requires an accurate
set of sensors and actuators. The reaction time of the actuators is critical
to achieve control, especially if one aims at the feedback cancellation of
nonlinear effects. The fact that some of the spatial growing modes have high
frequencies, leads to the suspicion that mechanical sensors and actuators,
even if highly miniaturized, will have an hard time to deal with the high
frequency instabilities that are known to be present in the transition.

Greater speed and flexibility would be achieved were it possible to act on
the flow by electromagnetic fields. With the possible exception of
electrolytes like seawater, a direct electromagnetic action on the
unmodified fluid\cite{Gailitis} \cite{Tsinober1} \cite{Moffat} \cite{Reed} 
\cite{Tsinober2} \cite{Henoch} \cite{Crawford} does not seem possible.
However, even for neutral fluids, improved control of the boundary layer
flow might be achieved by injecting in the leading edge of the airfoil a
stream of ionized gas, creating a thin ionized layer which might then be
acted upon by electromagnetic fields. In Ref.\cite{Vilela} a detailed
discussion is carried out of the effect of a streamwise directed electric
field on the velocity profile of an ionized boundary layer, taking into
consideration the fact that an injected stream of ionized gas leads to a
nonuniform charge profile. The study establishes reference values and design
estimates for the electric fields and ionization densities required for a
significant change of the velocity profile.

In the present paper a methodology is studied to assess the possibilities of
electromagnetic control of the TS\ precursor waves. Usually one thinks of
active control in terms of laminarizing the boundary layer flow. However the
opposite situation may also occur because, for example, in stalling prone
situations it might be useful to induce turbulence to avoid separation. Then
the fast reaction time of electromagnetic control might also be an asset.

\section{The stability equations}

Consider the Navier-Stokes equation 
\begin{equation}
\label{2.1}\frac{\partial \widetilde{U}}{\partial t}+(\widetilde{U}.\nabla )%
\widetilde{U}=-\frac 1{\widetilde{\rho }_m}\nabla \widetilde{p}+\widetilde{%
\nu }\triangle \widetilde{U}+\frac{\widetilde{\sigma }}{\widetilde{\rho }_m}%
\widetilde{E}+\frac{\widetilde{\sigma }}{c\widetilde{\rho }_m}\widetilde{U}%
\times \widetilde{B} 
\end{equation}
for an incompressible ionized fluid in an external electromagnetic field 
\begin{equation}
\label{2.2}\frac{\partial \widetilde{\rho }_m}{\partial t}+\nabla \cdot
\left( \widetilde{\rho }_m\widetilde{U}\right) =0 
\end{equation}
In orthogonal curvilinear coordinates, denote by $\left( \widetilde{u},%
\widetilde{v},\widetilde{w}\right) $ the streamwise, the normal and the
spanwise components of the physical velocity field $\widetilde{U}$. Define
also reference quantities and adimensional variables 
\begin{equation}
\label{2.3}
\begin{array}{llllll}
x=\frac{\widetilde{x}}{L_r}; & y=\frac{\widetilde{y}}{\delta _r}; & t=%
\widetilde{t}\frac{U_r}{L_r}; & u=\frac{\widetilde{u}}{U_r}; & v=\frac{%
\widetilde{v}L_r}{U_r\delta _r}; & w= 
\frac{\widetilde{w}}{U_r} \\ \rho _m=\frac{\widetilde{\rho }_m}{\rho _r}; & 
p=\frac{\widetilde{p}}{\rho _rU_r^2}; & \nu =\frac{\widetilde{\nu }}{\nu _r}%
; & \sigma =\frac{\widetilde{\sigma }}{\sigma _r}; & E=\frac{\widetilde{E}}{%
E_r} &  
\end{array}
\end{equation}
Typical values for the reference quantities, as used before\cite{Vilela},
are $U_r=100$ m s$^{-1}$, $L_r=1$ m, $\delta _r=10^{-3}$ m, $\rho _r=1.2$ Kg
m$^{-3}$, $E_r=500$ V cm$^{-1}$, $\sigma _r=15$ $\mu $C cm$^{-3}$, $\nu
_r=1.5\times 10^{-5}$ m$^2$ s$^{-1}$. Then R$_L=\frac{U_rL_r}{\nu _r}%
=6.66\times 10^6$ and $\frac 1{R_L}$ and $\frac{\delta _r^2}{L_r^2}=10^{-6}$
are small quantities.

Expressing (\ref{2.1}) in the adimensional variables (\ref{2.3}), assuming
that the product $k\delta $ of the airfoil curvature times the boundary
layer width is small and neglecting terms of order $R_L^{-1}$, $\frac{\delta
_r^2}{L_r^2}$ and $\frac{\widetilde{U}}c$ one is left with 
\begin{equation}
\label{2.4}
\begin{array}{l}
\frac{\partial u}{\partial t}+u\frac{\partial u}{\partial x}+v\frac{\partial
u}{\partial y}+w\frac{\partial w}{\partial z}=-\frac 1{\rho _m}\frac{%
\partial p}{\partial x}+\nu \omega \frac{\partial ^2u}{\partial y^2}+\frac
\gamma {\rho _m}\sigma E_x \\ \frac{\partial p}{\partial y}=\frac{\delta _r}{%
L_r}\gamma \sigma E_y \\ \frac{\partial w}{\partial t}+u\frac{\partial w}{%
\partial x}+v\frac{\partial w}{\partial y}+w\frac{\partial w}{\partial z}%
=-\frac 1{\rho _m}\frac{\partial p}{\partial z}+\nu \omega \frac{\partial ^2w%
}{\partial y^2}+\frac \gamma {\rho _m}\sigma E_z
\end{array}
\end{equation}
where $\gamma =\frac{L_r\sigma _rE_r}{U_r^2\rho _r}$ and $\omega =\frac{L_r^2%
}{\delta _r^2R_L}$ ($\gamma =62.5$ and $\omega =0.15$ for the reference
values above)

The aim is to study the stability of the steady state (laminar) solutions of
the above equations with regard to the precursor waves that develop in the
transition region. Therefore the variables are decomposed into steady state (%
$\overline{u},...$) and fluctuating components ($u^{^{\prime }},...$) 
\begin{equation}
\label{2.5}
\begin{array}{l}
u= 
\overline{u}+u^{^{\prime }} \\ v= 
\overline{v}+v^{^{\prime }} \\ w= 
\overline{w}+w^{^{\prime }} \\ p= 
\overline{p}+p^{^{\prime }} \\ E=\overline{E}+E^{^{\prime }} 
\end{array}
\end{equation}
and one looks for normal mode solutions of the form 
\begin{equation}
\label{2.6}\left\{ 
\begin{array}{c}
u^{^{\prime }} \\ 
v^{^{\prime }} \\ 
w^{^{\prime }} \\ 
p^{^{\prime }} 
\end{array}
\right\} =\left\{ 
\begin{array}{c}
\widehat{u}(y) \\ \widehat{v}(y) \\ \widehat{w}(y) \\ \widehat{p}(y) 
\end{array}
\right\} \exp \left\{ i\left( \alpha x+\beta z-\Omega t\right) \right\} 
\end{equation}
with a similar, but $y-$independent, form for the electric field

\begin{equation}
\label{2.7}E^{^{\prime }}=\widehat{E}\exp \left\{ i\left( \alpha x+\beta
z-\Omega t\right) \right\} 
\end{equation}
From the control point of view this implies the capability to have the
electric field react to the fluctuating velocity field with the same
frequency and wavelength, but eventually with some delay represented by the
phase of the complex amplitude $\widehat{E}$. To have this feedback
response, a distributed set of sensors should be available on the surface of
the airfoil. The sensors, of course, cannot measure the velocity field
itself but only some integrated effect, observable at the coordinate $y=0$
(see below).

In the transition region the quasiparallel hypothesis for the stationary
solution is a good approximation. Namely $\overline{v}=\overline{w}=\frac{%
\partial \overline{u}}{\partial x}=0$. This holds for example for one of the
scaling solutions in Ref.\cite{Vilela} 
\begin{equation}
\label{2.8}\overline{u}=u_e\left( 1-\exp \left( -y\frac \chi {\sqrt{u_e}%
}\right) \right) 
\end{equation}
for $\chi =\sqrt{\frac{\gamma \sigma _0\overline{E}_x}{\omega \nu \rho _m}}$%
\bigskip\ and a charge distribution profile 
\begin{equation}
\label{2.9}\sigma =\sigma _0\left( 1-\frac{\overline{u}}{u_e}\right) 
\end{equation}
It is the stability and controllability of this solution that is going to be
studied.

By differentiating Eqs.(\ref{2.4}) the pressure terms may be eliminated.
Then, keeping only the linear terms in the fluctuating fields and using (\ref
{2.6}) and (\ref{2.7}), one obtains 
\begin{equation}
\label{2.10}
\begin{array}{rcl}
\nu \omega \widehat{v}^{^{\prime \prime \prime \prime }}+i\theta \widehat{v}%
^{^{\prime \prime }} & = & i\alpha \left( 
\overline{u}^{^{\prime }}+\overline{u}^{^{\prime \prime }}\right) \widehat{v}%
+i\alpha \left( \overline{u}+\overline{u}^{^{\prime }}\right) \widehat{v}%
^{^{\prime }} \\  &  & +i\frac \gamma {\rho _m}\left\{ \sigma ^{^{\prime
}}\left( \alpha 
\widehat{E}_x+\beta \widehat{E}_z\right) -i\frac{\delta _r}{L_r}\sigma
\left( \alpha ^2+\beta ^2\right) \widehat{E}_y\right\}  \\ \nu \omega 
\widehat{w}^{^{\prime \prime \prime }}+i\theta \widehat{w}^{^{\prime }} & =
& i\alpha \overline{u}^{^{\prime }}\widehat{w}+i\alpha \overline{u}\widehat{w%
}^{^{\prime }}-\frac \gamma {\rho _m}\left\{ \sigma ^{^{\prime }}\widehat{E}%
_z-i\beta \frac{\delta _r}{L_r}\sigma \widehat{E}_y\right\} 
\end{array}
\end{equation}
with the boundary conditions 
\begin{equation}
\label{2.11}\widehat{v}(0)=\widehat{v}(\infty )=\widehat{w}(0)=\widehat{w}%
(\infty )=\widehat{v}^{^{\prime }}(0)=\widehat{v}^{^{\prime }}(\infty )=%
\widehat{w}^{^{\prime \prime }}(\infty )=0
\end{equation}
The last three boundary conditions are obtained from the continuity equation 
\begin{equation}
\label{2.12}i\alpha \widehat{u}+\widehat{v}^{^{\prime }}+i\beta \widehat{w}=0
\end{equation}
and the last equation in (\ref{2.4}).

In situations where the spanwise fluctuations may be neglected, the flow
becomes two-dimensional and a stream function may be defined for the waves 
\begin{equation}
\label{2.13}
\begin{array}{rcl}
u^{^{\prime }}=\frac{\partial \psi }{\partial y} & ; & v^{^{\prime }}=-
\frac{\partial \psi }{\partial x} \\ \psi  & = & \phi (y)\exp \left\{
i\left( \alpha x+\beta z-\Omega t\right) \right\} 
\end{array}
\end{equation}
Then the stability equation is 
\begin{equation}
\label{2.14}
\begin{array}{rll}
\nu \omega \phi ^{^{\prime \prime \prime \prime }}+i\theta \phi ^{^{\prime
\prime }} & = & i\alpha \left( \overline{u}\phi ^{^{\prime \prime }}+%
\overline{u}^{^{\prime \prime }}\phi \right) -\frac \gamma {\rho _m}\left\{
\sigma ^{^{\prime }}\widehat{E}_x-i\alpha \frac{\delta _r}{L_r}\sigma 
\widehat{E}_y\right\} 
\end{array}
\end{equation}
which is a simplified version of the Orr-Sommerfeld equation with a driving
term. The simplification arises from the fact that terms of order $1/R_L$
and $\delta _r^2/L_r^2$ have been neglected. In this form the equation may
be integrated once and reduced to a third order problem (see below).

\section{Stability and controllability results}

Consider first the spanwise stability of the scaling solution (\ref{2.8})
without control ($\widehat{E}_z=\widehat{E}_y=0$). The second equation in (%
\ref{2.10}) may be integrated once and the integration constant set to zero
using the boundary conditions (\ref{2.11}). Using the scaling solution (\ref
{2.8}) for $\overline{u}$ and changing coordinates to 
\begin{equation}
\label{3.1}\eta =1-\exp \left( -y\frac \chi {\sqrt{u_e}}\right) 
\end{equation}
one obtains 
\begin{equation}
\label{3.2}\left\{ (1-\eta )^2\frac{d^2}{d\eta ^2}-(1-\eta )\frac d{d\eta
}\right\} \widehat{w}+i\theta _1\widehat{w}=i\alpha _1\eta \widehat{w}
\end{equation}
where $\theta _1=\frac{\theta u_e}{\nu \omega \chi ^2}$ and $\alpha _1=\frac{%
\alpha u_e^2}{\nu \omega \chi ^2}$ , with boundary conditions 
\begin{equation}
\label{3.3}\widehat{w}(0)=\widehat{w}(1)=0
\end{equation}
Discretizing the $(0,1)$ interval, the calculation of the largest growing
modes becomes an algebraic generalized eigenvalue problem which is dealt
with by the QZ algorithm. In Fig.1a one plots the largest value of Re$%
(i\alpha _1)$ for real $\theta _1$ and in Fig.1b the largest value of Re$%
(-i\theta _1)$ for real $\alpha _1$. All modes having negative real parts,
the conclusion is that the scaling solution is both space- and time-
spanwise stable. Therefore we may take $\widehat{w}=0$ and use, for the
streamwise stability, the stream function stability equation (\ref{2.14}).

For the scaling solution (\ref{2.8}), with the same change of variables,
neglecting the term in $\widehat{E}_y$ because $\frac{\delta _r}{L_r}$ is a
small quantity, integrating once the equation and fixing the integration
constant with the boundary conditions, the result is the equation 
\begin{equation}
\label{3.4}
\begin{array}{c}
\left\{ (1-\eta )^3 
\frac{d^3}{d\eta ^3}-3(1-\eta )^2\frac{d^2}{d\eta ^2}+(1-\eta )\frac d{d\eta
}\right\} \phi +i\theta _1(1-\eta )\frac{d\phi }{d\eta } \\ =i\alpha
_1\left\{ \eta (1-\eta )\frac d{d\eta }-(1-\eta )\right\} \phi -\frac{\gamma
\sigma _0u_e^{3/2}}{\rho _m\nu \omega \chi ^3}(1-\eta )\widehat{E}_x 
\end{array}
\end{equation}
with boundary conditions 
\begin{equation}
\label{3.5}\phi (0)=\phi ^{^{\prime }}(0)=\phi (1)=0 
\end{equation}
Let first $\widehat{E}_x=0$ (uncontrolled equation). Using as before a
finite difference method and the QZ algorithm with the boundary conditions
imposed as three of the equations in the algebraic system, one obtains, for
the largest value of Re$(i\alpha _1)$ and real $\theta _1$, the results
shown in Fig.2. It means that there is a range of frequencies for which
there is spatial growth of the streamwise fluctuations. Therefore the
scaling solution is spatially unstable.

To derive a controlled equation one has to realize that the only physical
quantities, that it is reasonable to assume to be observable, are the
pressure fluctuations on the airfoil or the integrated effect of the
electrical current as seen at the surface of the airfoil. Pressure
fluctuations may be detected by a distributed set of microphones and the
integrated electrical current fluctuations are essentially the induced
magnetic field fluctuations on the spanwise direction. For definiteness I
will assume that a set of sensors is available to measure the effect of the
electrical current fluctuations. To achieve control this measurement is used
to modulate a variable component of the applied electric field. That is 
\begin{equation}
\label{3.6}\widehat{E}_x=k\int_0^\infty dy\sigma _0\left( 1-\frac{\overline{u%
}(y)}{u_e}\right) \widehat{u}(y)
\end{equation}
with $k$  a complex proportionality constant, the meaning of the phase being
the control delay. Then, the controlled equation is 
\begin{equation}
\label{3.7}
\begin{array}{c}
\left\{ (1-\eta )^3
\frac{d^3}{d\eta ^3}-3(1-\eta )^2\frac{d^2}{d\eta ^2}+(1-\eta )\frac d{d\eta
}\right\} \phi +i\theta _1(1-\eta )\frac{d\phi }{d\eta } \\ =i\alpha
_1\left\{ \eta (1-\eta )\frac d{d\eta }-(1-\eta )\right\} \phi -C(1-\eta
)\int_0^1d\eta \phi 
\end{array}
\end{equation}
with $C=\frac{k\gamma \sigma _0^2u_e^{3/2}}{\rho _m\nu \omega \chi ^3}$ .
The controllability problem amounts to find out whether all eigenvalues have
negative real parts in the integro-differential problem defined by Eq.(\ref
{3.7}). Let $\theta _1=60$ , the frequency for which the largest Re$(i\alpha
_1)$ is at its maximum. For real $C$, Fig.3 shows that for $C>1$ the largest
mode has spatial decay, hence the solution becomes stable. For the results
in Fig.4, let $C=\left| C\right| e^{i\varphi }$  with $\left| C\right| =1.5$
and variable phase $\varphi $. One sees that there is a range of phase
delays which stabilize the solution and, conversely, outside this range the
solution is strongly unstable.

\section{Conclusions}

1. Stabilization of a stationary configuration in an infinite-dimensional
system involves the study of infinitely many disturbance modes, some of
which may grow in time and space. In addition, the measurable observables,
to which some local control may react, involve the integrated effect of many
variables. Therefore the mathematical structure of the problem to be solved
is expected to be, as in this example, an integro-differential generalized
eigenvalue problem.

2. The unstabilizing disturbances that need to be controlled in extended
systems have in general a nontrivial space-time structure. Therefore a set
of distributed sensors and actuators is needed to achieve a space-time
controlling action.

3. The fact that, in practice, only global integrated variables are
observable, restricts the feedback control to these variables only.
Therefore, for extended systems, there is no guarantee that control will be
achieved in general and success is only to be expected in particular
favorable cases.

4. The laminar to turbulent transition, in the boundary layer, begins with
the appearance of downstream moving waves which at first grow slowly and may
be described by a linearized equation. After reaching a certain amplitude
however, the waves develop strong three-dimensional structures and
nonlinearities and a rapid transition to turbulence becomes unavoidable.
Therefore, if effective control is to be achieved, it is essential to have a
fast and locally accurate feedback to tame the instabilities before the
spanwise differential amplification of the TS waves begins to occur. Because
it is probably very difficult to obtain the required speed and accuracy with
mechanical microactuators, electromagnetic controlling schemes seem worth to
explore.

\section{Figure captions}

Fig.1 - (a) Space stability of the spanwise modes, (b) Time stability of the
spanwise modes

Fig.2 - Space instability of the streamwise scaling solution

Fig.3 - Largest Re$(i\alpha _1)$ for the controlled equation ($\theta _1=60$
and real $C$)

Fig.4 - Largest Re$(i\alpha _1)$ for the controlled equation ($\theta _1=60$%
, $\left| C\right| =1.5$ and variable phase)


\begin{thebibliography}{99}
\bibitem{Shinbrot}  T. Shinbrot; ''Progress in the control of chaos'',
Advances in Physics 44 (1995) 73.

\bibitem{Bushnell}  D. M. Bushnell and J. N. Hefner (Eds.); ''Viscous drag
reduction in boundary layers'', Progress in Astronautics and Aeronautics,
vol. 123, American Institute of Aeronautics and Astronautics, Washington
1990 and references therein.

\bibitem{Vilela}  R. Vilela Mendes and J. A. Dente; ''Boundary-layer control
by electric fields: A feasibility study'', Report no. physics/9705020. 

\bibitem{Joslin}  R. D. Joslin, G. Erlebacher and M. Y. Hussaini; ''Active
control of instabilities in laminar boundary layers - Overview and concept
validation'', Journal of Fluids Engineering 118 (1996) 494 and references
therein.

\bibitem{Schlichting}  H. Schlichting; ''Boundary-Layer Theory'' 6th.
edition, MacGraw Hill, New York 1968.

\bibitem{Young}  A. D. Young; ''Boundary Layers'', BSP\ Professional Books,
Blackwell, Oxford 1989.

\bibitem{Gailitis}  A. K. Gailitis and O. A. Lielausis; ''On the possibility
of drag reduction of a flat plate in an electrolyte'', Appl.
Magnetohydrodynamics, Trudy Inst. Fis. AN Latv. SSR 12 (1961) 143.

\bibitem{Tsinober1}  A. B. Tsinober and A. G. Shtern; ''Possibility of
increasing the flow stability in a boundary layer by means of crossed
electric and magnetic fields'', Magnetohydrodynamics 3 (1967) 103.

\bibitem{Moffat}  H. K. Moffat; ''On the suppression of turbulence by a
uniform magnetic field'', J. Fluid Mech. 28 (1967) 571.

\bibitem{Reed}  C. B. Reed and P. S. Lykoudis; ''The effect of a transverse
magnetic field on shear turbulence'', J. Fluid Mech. 89 (1978) 147.

\bibitem{Tsinober2}  A. Tsinober; ''MHD flow drag reduction'', in \cite
{Bushnell}, page 327.

\bibitem{Henoch}  C. Henoch and J. Stace; ''Experimental investigation of a
salt water turbulent boundary layer modified by an applied streamwise
magnetohydrodynamic body force'', Phys. Fluids 7 (1995) 1371.

\bibitem{Crawford}  C. H. Crawford and G. E. Karniadakis; ''Reynolds stress
analysis of EMHD-controlled wall turbulence. Part I. Streamwise forcing'',
Phys. Fluids 9 (1997) 788.
\end{thebibliography}
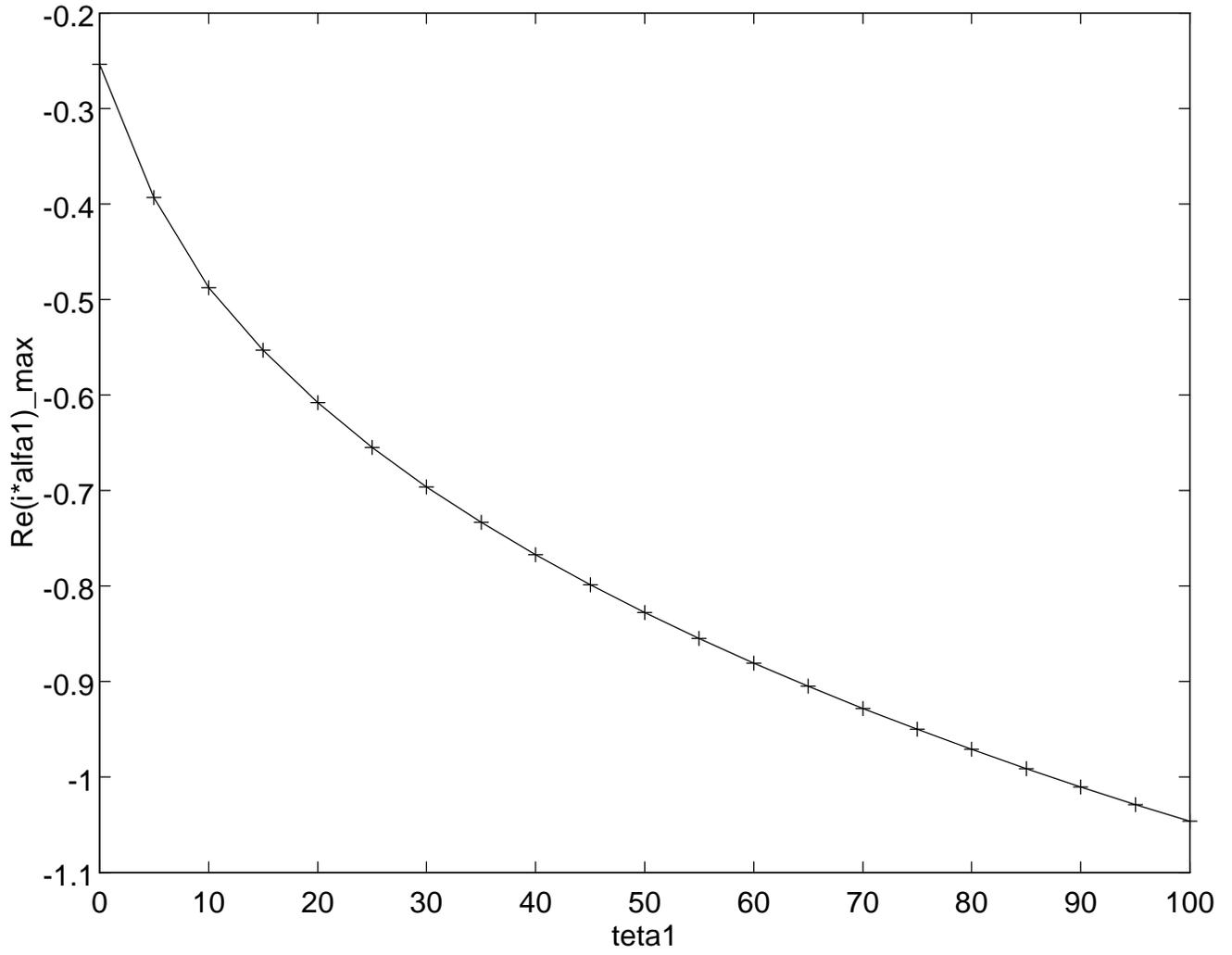
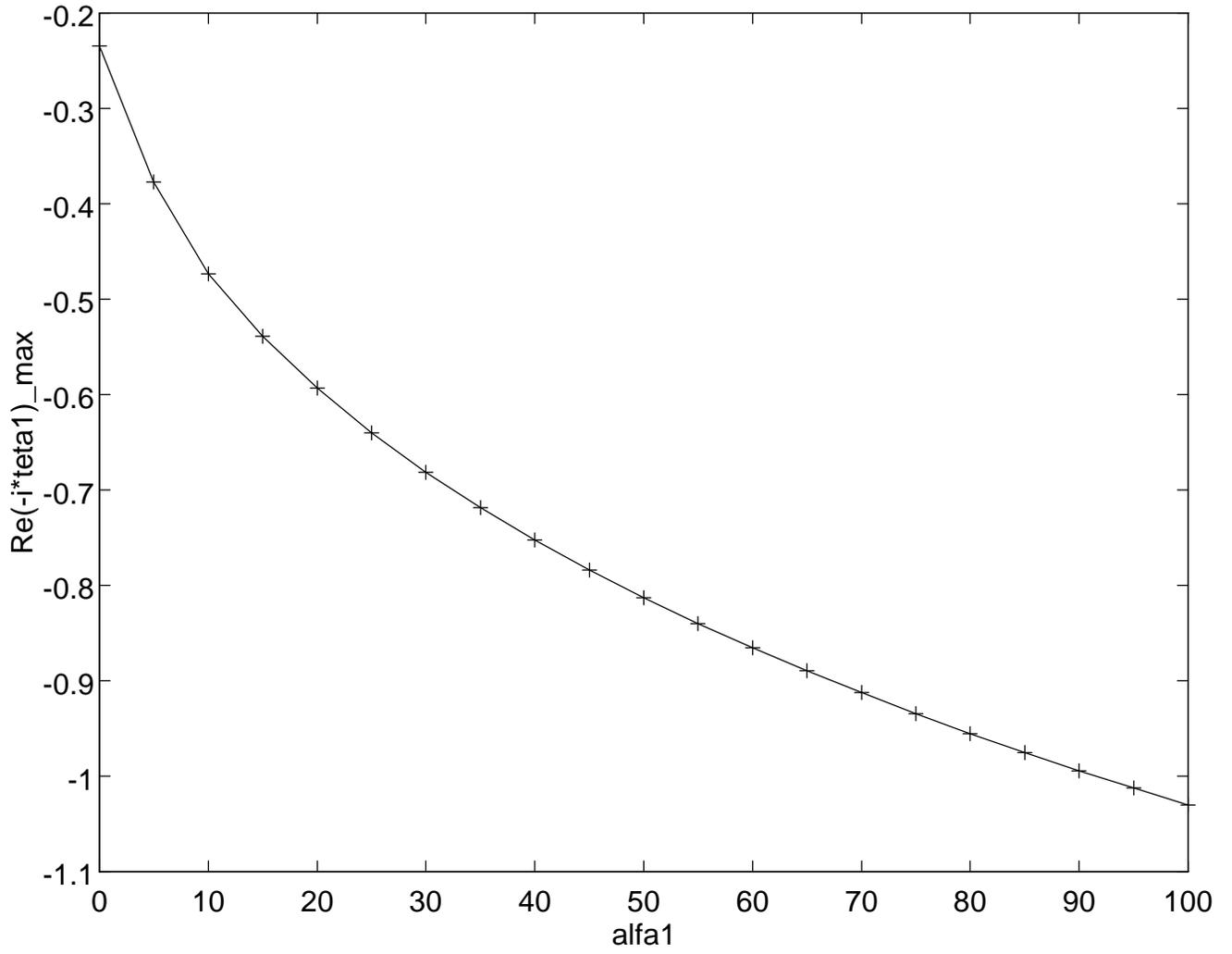
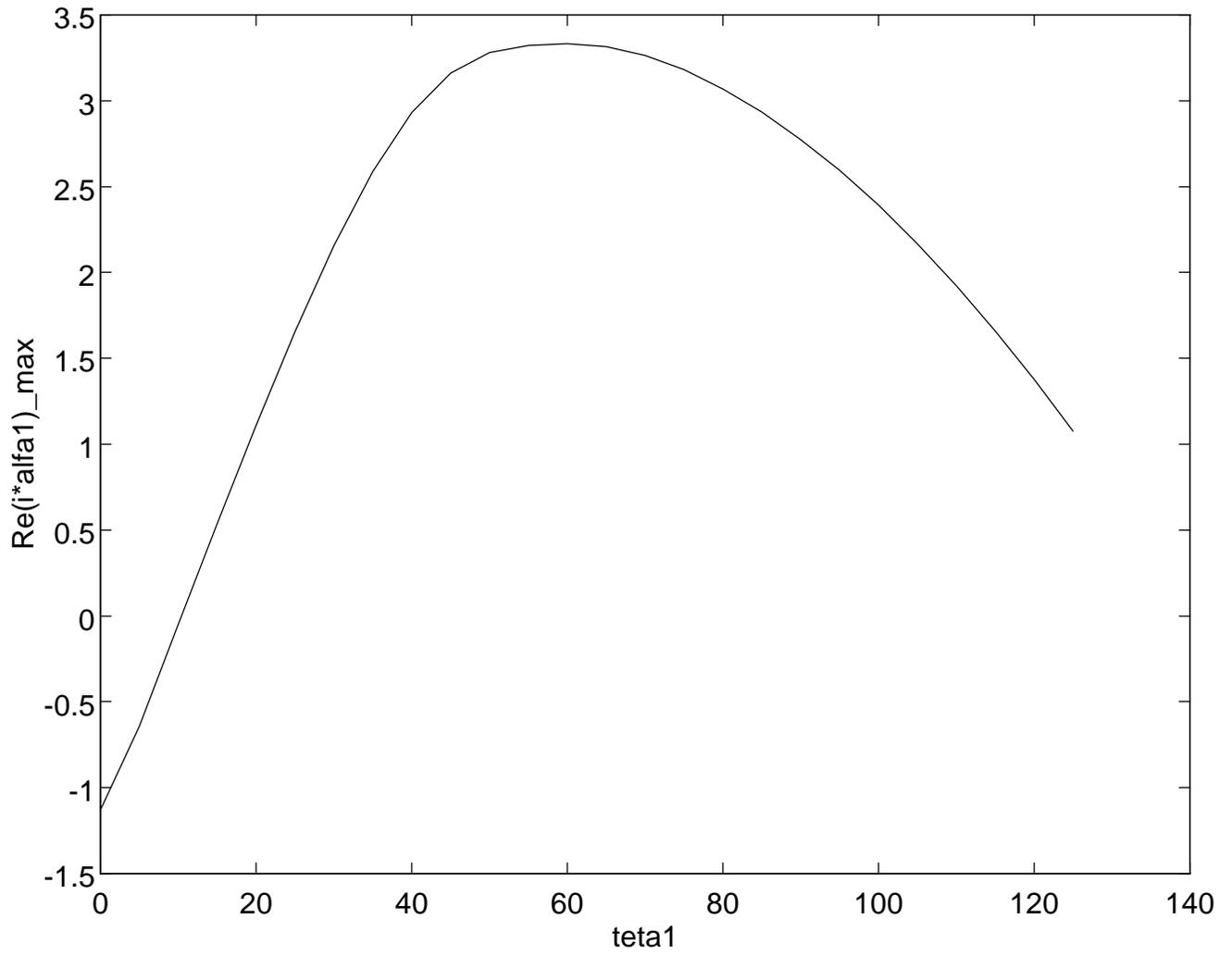
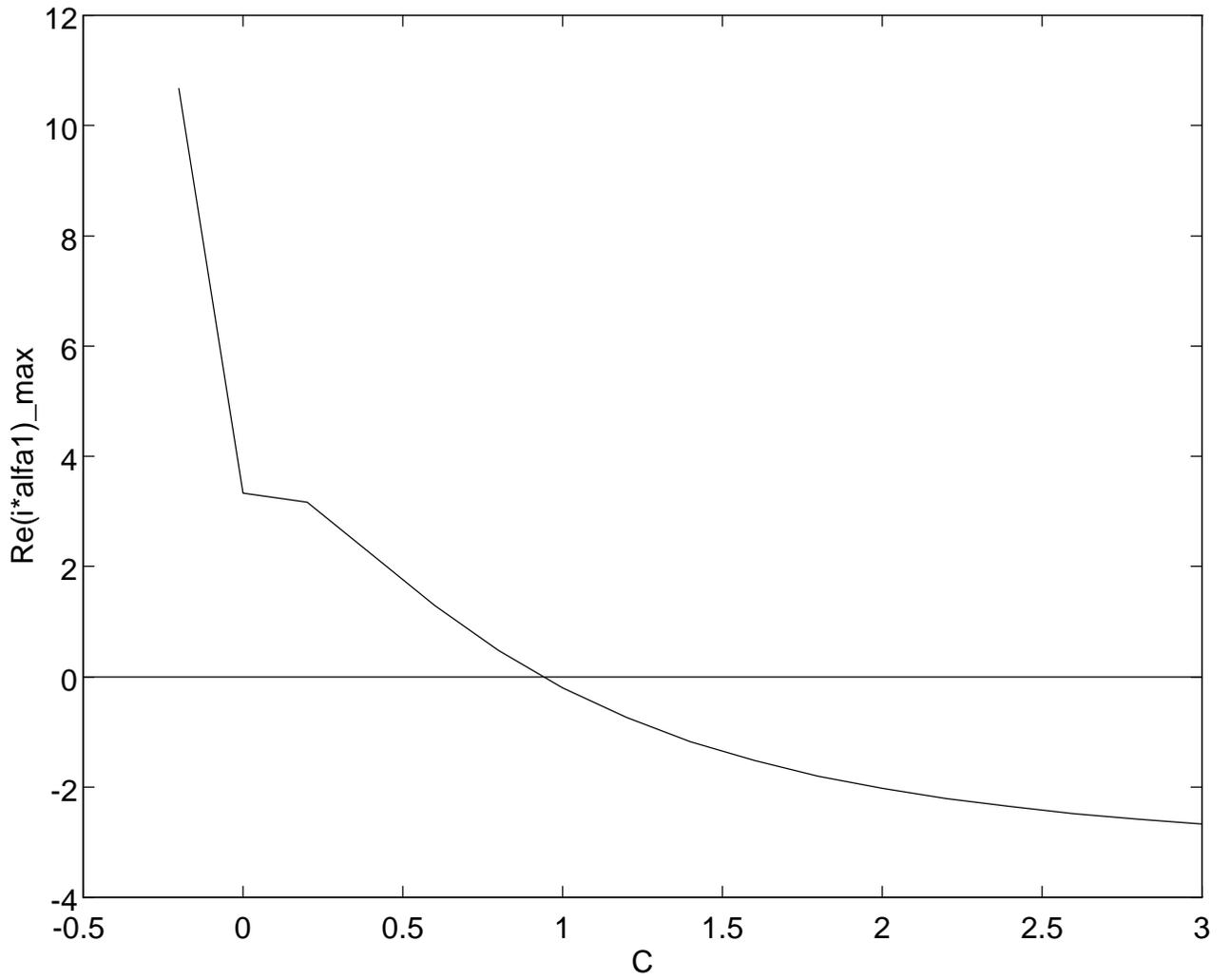
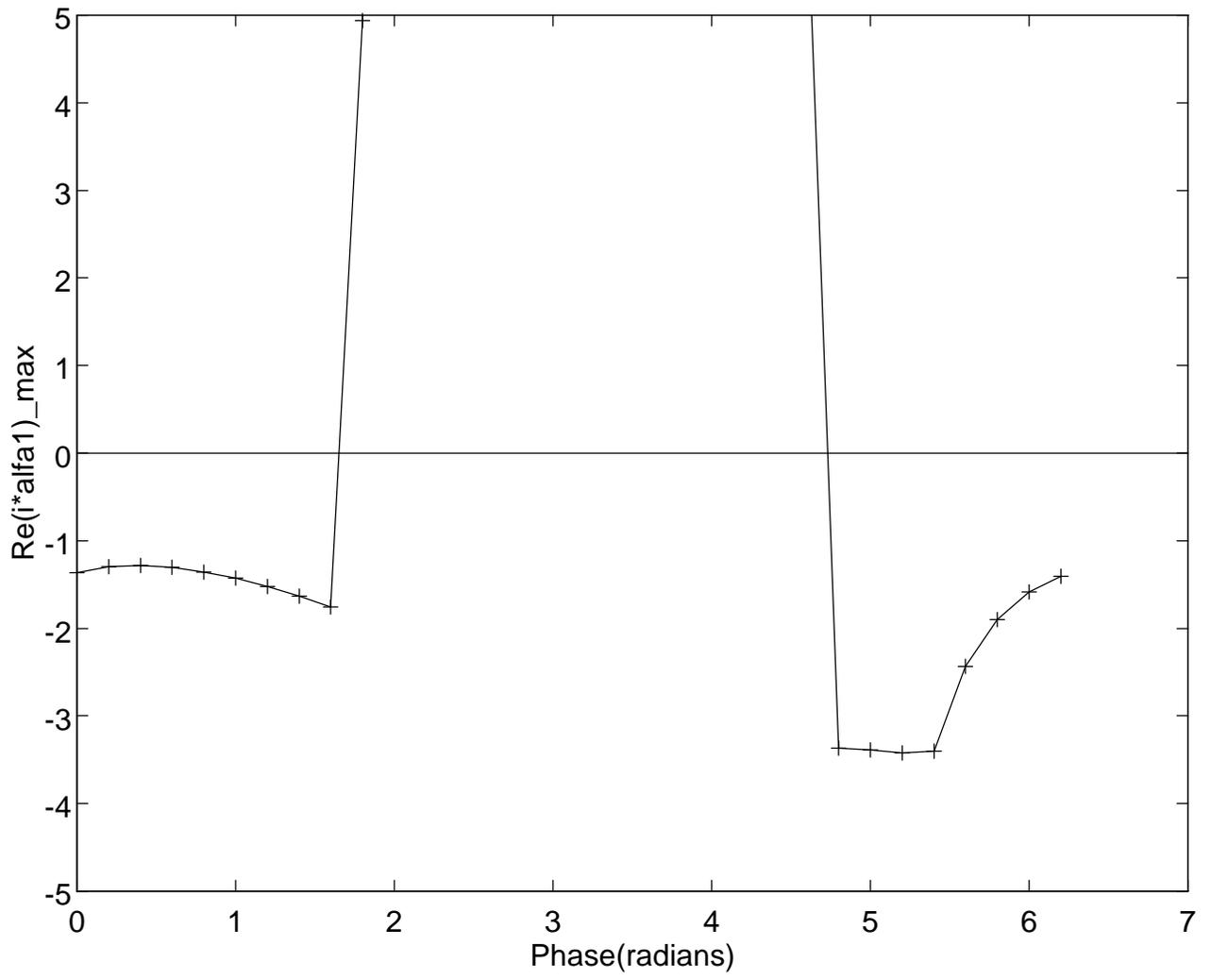
\end{document}